# Full counting statistics of quantum dot resonance fluorescence

Clemens Matthiesen[1]*, Megan J. Stanley[1]*, Maxime Hugues[2], Edmund Clarke[3], Mete Atatüre[1]

[1]Cavendish Laboratory, Department of Physics, University of Cambridge, JJ Thomson Avenue, CB3 0HE Cambridge, UK

[2] CNRS-CRHEA, rue Bernard Grégory, 06560 Valbonne, France

[3]EPSRC National Centre for III-V Technologies, University of Sheffield, Sheffield S1 3JD, UK

*equally contributing authors

**The electronic energy levels and optical transitions of a semiconductor quantum dot are subject to dynamics within the solid-state environment. In particular, fluctuating electric fields due to nearby charge traps or other quantum dots shift the transition frequencies via the Stark effect. The environment dynamics are mapped directly onto the fluorescence under resonant excitation and diminish the prospects of quantum dots as sources of indistinguishable photons in optical quantum computing. Here, we present an analysis of resonance fluorescence fluctuations based on photon counting statistics which captures the underlying time-averaged electric field fluctuations of the local environment. The measurement protocol avoids dynamic feedback on the electric environment and the dynamics of the quantum dot's nuclear spin bath by virtue of its resonant nature and by keeping experimental control parameters such as excitation frequency and external fields constant throughout. The method introduced here is experimentally undemanding.**

Introduction

Recent research on the optical properties of single semiconductor InAs/GaAs quantum dots (QDs) explored quantum effects well known from atomic optics, such as photon antibunching [1,2] or the dressing of the energy levels by a resonant light field [3,4]. Experimental evidence of the near-ideal behaviour of QD emission under resonant excitation was reported in spectral measurements [5] and by interference of two photons emitted by the same QD [6, 7]. However, while the observed optical coherence confirms the absence of a significant fast dephasing mechanism, other measurements present clear deviations from the theoretical predictions, such as broadened absorption linewidths [4,8-13] and narrower than expected antibunching signatures in the intensity autocorrelation. Similar effects are observed in coherent control schemes such as coherent population trapping [14-16]. An explanation lies in consideration of the solid-state environment of the QD, e.g. the nuclear Overhauser [17] and electric field fluctuations.

The Overhauser field originates from the hyperfine interaction of the QD's nuclear spin bath with a resident electron spin and is a property inherent to InAs/GaAs QDs. The nuclear field fluctuations show a consistent behaviour for these types of QDs, typically amounting to an effective magnetic field of 20-30 mT [17]. Electric field noise is caused by the charge dynamics in doped layers, semiconductor interfaces, or by metastable trapped charges in defects and other QDs nearby. The net time-dependent electric field produces a proportional shift in the QD resonance via the Stark effect [18]. Particular sample circumstances and models of defect dynamics have been subject to a number of recent studies [10, 13, 19-21], where timescales and magnitudes of spectral diffusion vary greatly between different structures and samples. A robust and experimentally undemanding method to measure steady-state spectral diffusion of individual QDs can therefore be a powerful tool to characterise the QD sample quality.

Here we take advantage of the frequency selectivity of single QD resonance fluorescence for precise and minimally invasive measurements of spectral diffusion. Using continuous-wave excitation and collection of the QD's fluorescence in a confocal microscope [5] we record fluorescence intensity time traces. Assigning the variable $k_{bin}$ to the number of photons detected in a time bin we compute the probability $P(k_{bin})$ to obtain exactly $k_{bin}$ photon counts, and hence acquire the photon counting histogram for the time trace. In the following we develop a full counting statistics model showing that the histogram shape is a very sensitive probe of spectral diffusion, and find remarkable agreement with the experimental data. Fitting the full counting model to experimental histograms we can quantify spectral diffusion effects with a precision better than 0.05 µeV.

Results

Photons scattered by a single emitter display sub-Poissonian intensity statistics, resulting in the well-known antibunching dip in the intensity autocorrelation [22] on the timescale of the radiative lifetime $T_1$. On longer timescales photons from an ideal two-level system simply follow Poisson statistics characterised by a mean rate of photon emission and a variance equal to the mean. For a semiconductor QD, the presence of electric field noise in the environment results in a time-dependent detuning between the resonant excitation laser and the QD's instantaneous resonance frequency, therefore modifying the photon emission rate. This concept is illustrated in Figure 1. In panel (a) the QD absorption lineshape is shown for three distinct values of the local electric field. When a laser of fixed frequency and power drives the QD transition, the mean fluorescence intensity is determined by the detuning of the transition from the laser due to the environment-induced Stark shift in each case. The photon counting intensity histograms arising from these three situations are displayed in the inset to the right using the same colour code. For a mean number of $m$ photons per measurement bin, the probability distribution of $k_{bin}$ detections per bin follows the Poisson distribution

$$P(k_{bin}) = \frac{m^{k_{bin}} e^{-m}}{k_{bin}!} \tag{1}$$

The mean *m* is linked to the QD fluorescence through the collection and detection efficiency $\eta_{detection}$, the duration of a detection bin $t_{bin}$ and the excited state population:

$$m(s,\Delta) = \eta_{detection} \times t_{bin} \times \frac{1}{2T_1} \frac{s}{1+s+2(2\pi\Delta)^2 T_1 T_2}, \tag{2}$$

where $\Delta$ is the detuning from resonance and $T_2$ is the coherence time of the two-level system. The saturation parameter *s* is linked to the Rabi frequency by $s = \Omega^2 T_1 T_2$. As a shorthand notation we define the amplitude *a(s)* to denote the mean counts on resonance $a(s) = m(s, \Delta = 0)$, providing a simple link to experimental data.

If, for simplicity, we assume the QD experiences the three local field strengths from the concept sketch in Fig. 1(a) for the same amount of time the cumulative photon counting histogram will be the sum of the individual histograms. This sum is plotted as grey shaded curve. The resulting cumulative histogram shape clearly differs from its Poissonian components.

In order to produce a more realistic model now we describe the time-dependent detuning due to electric field noise by a continuous probability distribution. For sufficiently long measurement times the probability distribution is fully sampled. The description of the photon counting histograms $P(k_{bin})$ from equation (1) is then modified to

$$P(k_{bin}) = \sum_\Delta W(\Delta) \times \frac{m(s,\Delta)^{k_{bin}} e^{-m(s,\Delta)}}{k_{bin}!}. \tag{3}$$

The weighting function $W(\Delta)$ describes the probability of measuring a particular environmentally induced frequency shift in an instantaneous measurement, and we assume here that it follows a Gaussian distribution for long measurement times:

$$W(\Delta) = \exp\left[-\frac{1}{2}\left(\frac{\Delta - \delta_{\text{average}}}{\Delta_{\text{Diffusion}}}\right)^2 8\ln 2\right]. \tag{4}$$

A Gaussian distribution of frequency shifts is a reasonable assumption for a linear Stark coefficient and electric field noise, where a large number of fluctuating electric field sources are expected to contribute and all configurations may be explored over long times. The Gaussian is centred on $\delta_{\text{average}}$ to describe the effect of spectral diffusion when setting the laser at a finite average detuning from the QD resonance. The magnitude of the spectral diffusion is described by the full-width-at-half-maximum $\Delta_{\text{Diffusion}}$ of the Gaussian distribution.

Using equations (3,4) we calculate and plot $P(k_{\text{bin}})$ in Figure 1(b) for a range of laser detunings $\delta_{\text{average}}$ with physically reasonable parameters: $T_1$ = 0.65 ns, $\Delta_{\text{Diffusion}}$ = 250 MHz, amplitude $a$ = 100 counts per bin. To the right, three histograms are reproduced in the conventional form for specific detunings: 220 MHz (blue), 120 MHz (green), 40 MHz (red). Histograms can be double-peaked (e.g. around 120 MHz detuning here) and may generally appear to be bi-modal, but we note this is a natural consequence of taking a sum of Poisson distributions. Small count rates naturally give rise to histograms with small variance, while high count rates are associated with broad distributions. At intermediate detuning the mean intensity is most sensitive to a shift in the instantaneous resonance frequency and thus a more equal weighting is given to both high and low mean Poisson distributions.

The model predicts a unique histogram for each combination of parameters ($\Delta_{\text{Diffusion}}$, $\delta_{\text{average}}$, $a$) and we establish next how the model relates to experimental data (for details on the samples please refer to the Methods section). First, we measure the QD transition lifetime $T_1$ via pulsed resonant excitation. Additionally we find the saturation parameter $s$ by a power dependent measurement of the emission compared to Eq. (2). Records of time-resolved resonance fluorescence containing on the order of $10^7$ detection events are used to obtain low noise histograms. This corresponds to acquisition times in the range of 10-100 seconds in our setup. The fluorescence time traces are binned to a resolution of 0.1-1 ms. Here, the acquisition time naturally sets the slowest timescale of fluctuations to which the measurement is sensitive, while the binning time limits the bandwidth for faster fluctuations.

Figure 2(a) displays a histogram of resonance fluorescence, data as filled circles, from the negatively charged trion transition of QD A in a semi-logarithmic plot. Error bars represent a statistical uncertainty of one standard deviation ($\sigma$). A least-squares fit to the model is shown as continuous blue curve and the residual is given in the lower panel. Additionally, the red line displays the Poissonian histogram expected in the absence of spectral diffusion with identical detuning and saturation parameters. The inset shows the data and the fit on a linear scale. In Figure 2(b) we calculate the square of the deviations between the model and data for a three-dimensional grid of the three model parameters, detuning, diffusion width and amplitude. In this 3D space a single minimum exists, plane cuts through which are projected to the sidewalls of the plot. This single minimum confirms that the experimental data can also be associated with a unique set of parameters.

It is necessary to check that the parameters obtained are physically reasonable. Exemplary data are shown for QD B and QD C in Figure 3. Panels (a) and (c) display the normalised histogram data and fits to the model for constant excitation power and four detuning values in both cases. The transition probed is indicated in the top left corner of the panels. The skewness of the histograms is well captured by the model. In particular, close to resonance (see red data curves) the histograms exhibit a tail on the left of the maximum, i.e. at lower counts, while the tail appears on the right for large detuning (see violet data curves). In Figure 3(b) and 3(d) the behaviour of two fit parameters is summarised by plotting the Gaussian diffusion $\Delta_{\text{Diffusion}}$ width versus the detuning $\delta_{\text{average}}$. The error bars indicate a doubling of the error in the fit. The magnitude of spectral diffusion shows little spread or systematic dependence on the detuning, which is consistent with the reasonable expectation that environment noise is independent of the rate of excited state population at constant incident laser power. For other QDs on the same Schottky diode sample as QDs B and C we measured diffusion widths $\Delta_{\text{Diffusion}}$ ranging from 50-150 MHz. The sample of QD A (cf. Fig. 2(a)) was operated in a regime where a small current is flowing in the diode. Diffusion widths of 300 MHz to 1 GHz were observed, indicating a strong enhancement of electric field noise. In terms of the electric field strength this corresponds to 0.05-0.15 kV/cm.

The discussion of spectral diffusion has focussed so far on continuous spectral shifts which can be described by a single (Gaussian) probability distribution. Discontinuous spectral jumps and 'blinking' on the other hand are also observed [23, 24], often for samples with a large number of defects or high QD density, and commonly in ungated devices. Blinking denotes the temporary inability of an emitter to scatter photons and is a major issue for colloidal QDs [25]. In contrast, a spectral jump shifts the emitter's resonance with respect to a resonant laser, resulting in a modified scattering rate. We demonstrate that counting statistics provides a powerful method to identify these effects through fluorescence fluctuations. Figure 4 presents histogram data for the $X^{1-}$ transition of QD D (blue curve). Applying the model with a single Gaussian probability distribution for the detuning (cf. equation (4)) yields good agreement to part of the data, but only if the area of the fit is reduced from unity. This fit is shown as red shaded area in the main figure. In the panel below, the residual to the data (blue diamonds) highlights a second mode of the detuning probability distribution is needed to obtain a fit for the full data set. The histogram due to the second mode is shown as red line with blue shading in the residuals plot. On the right-hand side of Fig. 4 the physical process is explained in a sketch. The local electric field switches discontinuously between two values, consequently the QD resonance frequency jumps between two values. The exciting laser is blue-detuned with respect to the two resonance frequencies here. The mode closer to the laser is populated with high probability (96-97%). A spectral jump into the second mode reduces the fluorescence intensity; it shows up on the histogram at lower counts. By tuning the excitation laser to lower frequencies we observe the dominant mode moving through a maximum in intensity before weakening. At these lower frequencies a spectral jump brings the QD closer to resonance with the excitation, such that we observe the less likely second mode at higher counts in the histogram.

Discussion

Photon counting statistics have long been part of the fluorescence fluctuation spectroscopy toolbox in the bio-physical sciences and chemistry [26-28]. In a particular experiment fluorescence from several molecular species may be collected at the same time or single molecules could be diffusing in and out of the excitation volume, leaving fingerprints of the dynamics in the intensity statistics of the fluorescence. In contrast, we employ the resonance fluorescence of a single stationary emitter which is sensitive to the local fluctuations of electric and magnetic fields. We introduced a simple model of Gaussian-weighted diffusion of the resonance frequency for a Lorentzian absorption lineshape which captures the experimentally observed photon counting statistics. We note that previous work has used a Lorentzian probability distribution [16] for spectral diffusion, which does not give good results in our case. The long tails of the Lorentzian give rise to Poissonian components with small means and consequently small variance, significantly overestimating the histogram amplitude at low counts.

Intrinsic to all InGaAs dots is the bath of nuclear spins; in the region of the carrier wavefunctions this acts to produce a fluctuating effective magnetic field, the Overhauser field [17]. The interaction of the electron ground state with this field broadens the absorption lineshape and also leads to a deviation from the typical Lorentzian form. In particular, the slope around the resonance is reduced, and thus intensity fluctuations due to electric field noise on top of the nuclear spin dispersion are smaller around resonance compared to larger detuning. This effect becomes apparent when the model is applied to charged exciton transitions ($X^{1-}$). Consistently smaller diffusion widths are recovered close to resonance in these cases. Figure 3(d) displays this phenomenon. Incorporating the Overhauser field fluctuations in addition to the electric field fluctuations in this model would give more accurate physical parameters in situations where the nuclear spin dynamics play an important role. However, the simple Gaussian diffusion model can reproduce intensity fluctuations correctly. This is primarily due to the dominance of electric field noise for the timing resolution of the present measurements.

This measurement technique gives reliable results for moderate sample sizes ($10^7$ detection events, corresponding to ~ 10 s acquisition time in our setup), and the required timing resolution (on the order of 100 µs) is easily accessible with affordable data acquisition cards or dedicated counters. In contrast to conventional direct spectral measurements of spectral diffusion in solid-state systems [29], we utilise the frequency selectivity of resonance fluorescence to quantify spectral diffusion with a resolution exceeding 0.05 µeV in a simple fluorescence intensity measurement. The high sensitivity provides access to spectral diffusion processes far below the resolution limit of typical spectrometers.

The simplicity of the model makes it an attractive method to characterize the spectral diffusion of a sample related to the electric field. It is the properties of the electric field noise that differ widely between samples and sensitively depend on the sample structure and growth conditions. In this context, full counting statistics are a powerful method to provide information on electric field noise for different sample circumstances.

In summary, we advocate full counting statistics of resonance fluorescence as a method to characterise noise due to the local electric environment, and the resulting photon quality. This supplements existing techniques such as spectral or first-order correlation measurements, and intensity correlations. In combination with these techniques a more complete understanding of environment dynamics can be gained. As a direct next step, we plan to use this method to evaluate the effectiveness of active feedback on the environment.

Methods

Results presented in this report stem from two InAs QD samples fabricated into Schottky diodes, both originating from the same wafer. Measurements of other QD wafers were conducted, yielding similar results. With regards to the particular data included here, measurements on QDs A, C and D were conducted in a helium-flow cryostat at a sample temperature of 20 K, while QD B was measured at 4 K in a liquid helium bath cryostat. For both samples measurements were taken on QDs located within an area of $\sim$ 30x30 $\mu m^2$. The QD wafer was grown by molecular beam epitaxy and contains a single layer of self-assembled InAs/GaAs QDs embedded in a Schottky diode for charge state control. A GaAs tunnel barrier of nominally 35 nm separates an n-doped ohmic contact layer from the QDs, on top of which there are, in order, a 10 nm GaAs cap, a 50 nm AlGaAs tunnel barrier and another GaAs cap of 90 nm thickness. 20 pairs of GaAs/AlGaAs layers form a distributed Bragg reflector (DBR) below the QD diode. The DBR and sample surface form a weak cavity, enhancing light outcoupling around wavelengths of 980 nm. With a super-hemispherical zirconia solid immersion lens [30] placed on top of the sample we achieve an outcoupling efficiency of ~15 % at those wavelengths. Optical addressing of single QDs is achieved using confocal microscopy in both a liquid helium flow cryostat and a bath cryostat using power- and frequency-stabilised external cavity diode lasers. Scattering of the resonant laser light is suppressed using crossed linear polarisers in excitation and detection paths. The technique was introduced in Ref. [3] and characterised in Ref. [5]. QD resonance fluorescence is coupled into single mode fibres and detected with single photon counting avalanche photodiodes. Count rates between 0.5-3 MHz were registered in the experiment at high Rabi frequencies, depending on the setup (flow or bath cryostat), and the QD wavelength.

Fits of the model to the data were obtained in a least squares fit routine programmed in Matlab.

Acknowledgments

This work was supported by the University of Cambridge. CM acknowledges Clare College, Cambridge, for financial support through a Junior Research Fellowship. The authors acknowledge fruitful discussions with J. Hansom, C. Le Gall, R. Stockill and C. H. H. Schulte.

Author contribution statement

C.M. and M.J.S. developed the model, took the measurements and analysed the data. C.M., M.J.S and M.A. wrote the paper. M.H. and E.C. grew the sample, C.M. processed the sample into a Schottky diode.

The authors declare no competing financial interests.

Correspondence and requests for materials: cm467@cam.ac.uk

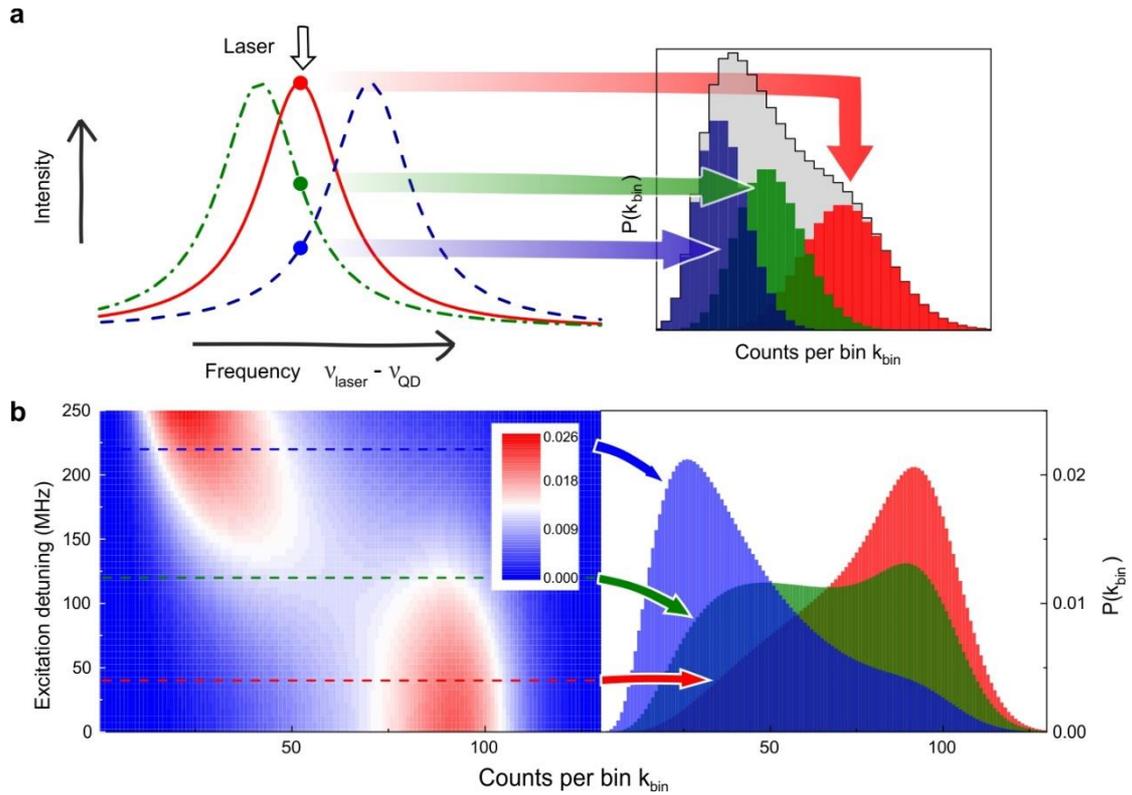

**Figure 1: Influence of spectral diffusion on resonant photon counting statistics.** (a) The QD optical transition frequency shifts as a consequence of local electric field noise such that a laser of fixed frequency drives the transition with a time-dependent detuning. Here we show the absorption lineshape at three instances in time. Right panel: intensity histograms corresponding to the three detunings depicted on the left (same colour code). The cumulative histogram (grey shaded) strongly deviates from the ideal Poissonian distribution. (b) Calculation of full counting statistics histograms for a Gaussian diffusion distribution with a full-width-half-maximum $\Delta_{\text{Diffusion}}$ equal to the natural linewidth of 250 MHz. The excitation detuning $\delta_{\text{average}}$ denotes the average detuning of the laser from the QD resonance. The saturation parameter is $s=1$, the average count rate on resonance $a$ would correspond to 100 counts per bin in the absence of environment effects. Right panel: Linecuts from the calculation for three detunings.

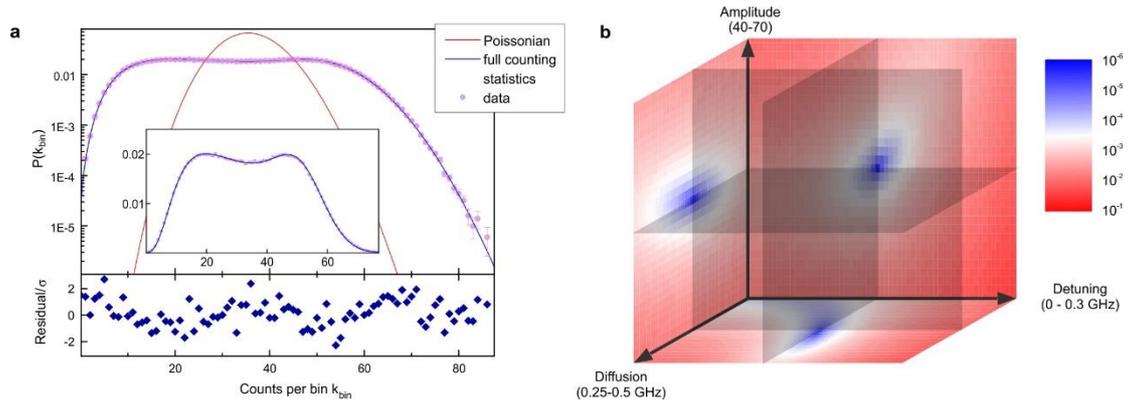

**Figure 2: Fitting quantum dot resonance fluorescence histograms using full counting statistics.** (a) Photon counting histogram of resonance fluorescence intensity for the charged exciton transition ($X^{1-}$) of QD A at finite detuning, s=1.7, acquisition time of 50 s and bin size of 100 µs. Data is shown as filled circles, a least squares fit according the model as blue line. The red line maps out the expected histogram at the same detuning, but in the absence of spectral diffusion. Fit parameters: amplitude a = 55.3, detuning $\delta_{average}$ = 148 MHz, Gaussian diffusion width $\Delta_{Diffusion}$ = 364 MHz. Inset: same data and fit on linear scale. Bottom: residual of fit.
(b) Calculation of least squares error using the data from panel (a) in the parameter space of amplitude, detuning and diffusion. The sidewalls show projections of plane cuts through the global error minimum. The actual location of the planes in the parameter space is indicated by the shaded squares. The model error has a single well-defined minimum.

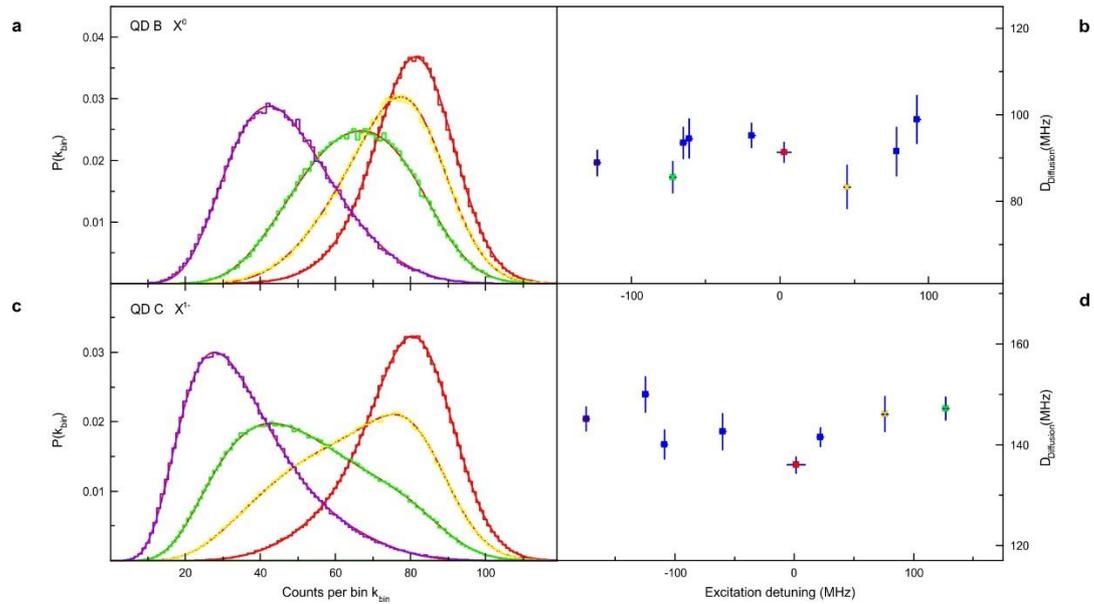

**Figure 3: Signatures of electric field fluctuations in photon counting histograms at resonance and for detuning.** (a), (c) Photon counting histograms for QD B and QD C at fixed excitation power and a range of detunings fitted using full counting statistics. The data follow a colour scale where red goes to violet as detuning increases. Fits are shown as red continuous curves on top of the data. Experimental parameters: (a) s=0.8, 70 s acquisition time, 500 μs bins, (c) s=1.2, 100 s acquisition time, 200 μs bins. (b), (d) Gaussian diffusion width as calculated from the model as a function of excitation detuning, also calculated from the model. The coloured data points show model parameters for the histograms (a, c) of the same colour. The diffusion varies little with detuning, pointing to local intrinsic electric-field fluctuations. Error bars represent a two-fold increase from the least-squares error.

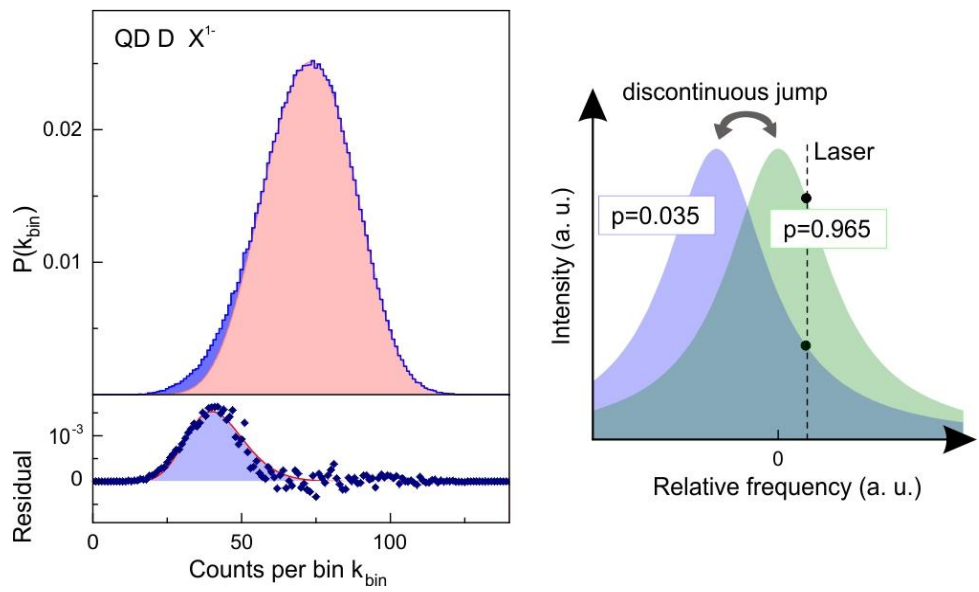

**Figure 4: Identifying spectral jumps through the photon statistics.** (a) Intensity histogram for the $X^{1-}$ transition of QD D ($s$=0.8, $\delta_{average}$ =70MHz, $\Delta_{Diffusion}$ =70MHz). Applying full counting statistics reveals a bimodal detuning probability distribution. The data is shown as blue stepped curve; the red curve with red shading shows a fit to the dominant of the distribution modes. Bottom, the residual (data as blue diamonds) reveals the second weaker mode. (b) Sketch of the physical process. The QD resonance switches discontinuously between two frequencies. Relative frequency denotes the detuning between laser and the QD resonance in the dominant distribution mode. The fit from (a) puts the probability to be in the brighter (darker) mode to ~0.965 (0.035).